\documentclass[aps,pre,twocolumn]{revtex4-1}
\usepackage{bm, epsfig}
\usepackage{amsmath,amssymb}
\usepackage{graphicx, subfigure}

\begin{document}

\author{Sourav Karmakar and Srihari Keshavamurthy}
\email{srihari@iitk.ac.in}
\affiliation{Department of Chemistry, Indian Institute of Technology,
Kanpur, Uttar Pradesh 208 016, India}

\title[]{Relevance of the resonance junctions on the Arnold web to dynamical tunneling and eigenstate delocalization}

\begin{abstract}
In this work we study the competition and correspondence between the classical and quantum routes to intramolecular vibrational energy redistribution (IVR) in a three degrees of freedom model effective Hamiltonian. Specifically,
we focus on the classical and the quantum dynamics near the resonance junctions on the Arnold web that are formed by intersection of independent resonances. The regime of interest models the IVR dynamics from highly  excited initial states  near  dissociation 
thresholds of  molecular systems wherein both classical and purely quantum, involving dynamical tunneling,  routes to IVR coexist. In the vicinity of a resonance junction classical chaos is inevitably present and hence one expects 
the quantum IVR pathways to have a strong classical component as well. We show that with increasing resonant coupling strengths the classical component of IVR leads to a transition from coherent dynamical tunneling to incoherent dynamical 
tunneling. Furthermore, we establish that the quantum IVR dynamics can be predicted based on the structures on the classical Arnold web. In addition, we investigate the nature of the highly excited eigenstates in order to identify the 
quantum signatures of the multiplicity-$2$ junctions. For the parameter regimes studies herein, by projecting the eigenstates onto the Arnold web, we find that eigenstates in the vicinity of the junctions are primarily delocalized 
due to dynamical tunneling. 
\end{abstract}
\maketitle

\section{Introduction} 
\label{intro}

Uncovering the mechanisms, classical and quantum, of intramolecular vibrational energy redistribution (IVR)\cite{uzer91,nesbittfield96,mgrub00,mgrubpwol04,ksriacp13,Leitner2015} in an isolated molecule continues to be an area of significant 
interest due to several reasons. Firstly, IVR lies at the heart of all statistical rate theories\cite{BaerHasebook} with the assumption that the timescale for IVR is short in comparison with a typical timescale for a reaction. Secondly, incomplete IVR on chemically significant timescales raises the hope for mode-specific chemistry\cite{Crim2008,huaguo15} 
and the possibility of active control\cite{Leeetal2012} of reactive events. Thirdly, interpreting  overtone spectroscopy\cite{Kellman2007,HermanPerry2013} in terms of emergent, qualitatively new, motions at high excitation energies requires  precise mechanistic insights into 
the energy flow dynamics. From a fundamental perspective,  thermalization of an initial hot spot in the molecule i.e., IVR, is connected to topics of significant recent interest to the condensed matter community - eigenstate thermalization hypothesis (ETH)\cite{rigol2008,Eisert2015,alessio2016} and many body localization (MBL)\cite{NandHuse2015}. There is little doubt that insights into IVR, or ETH and MBL for that matter, will have far 
ranging impact on issues ranging from identifying correct reaction mechanisms in gas\cite{Carpenter2013} and solution phase\cite{Carpenteretal2016} to the nature of heat flow in proteins\cite{Helbing2012,leitner2008} and the efficiency of nano-junctions in molecular 
electronic devices\cite{Segal2016,PandeyLeitner2016}. 

Insights into several aspects of the IVR dynamics have been obtained over the past few decades with novel experimental\cite{Hamilton1986,Pate2008,Kurochkin2007,Ferrante2016} and theoretical approaches\cite{mgrub00,ksriacp13,Leitner2015,mehta1995,Burin2010,ShiGeva2003,JainSubotnik2018,mgrub99,Chuntonov2017,Fujisaki2017,Meyer2006}. 
These studies have brought out the utility of approximately conserved quantities called polyads, emphasized the importance of local density of states coupled to the initial state of interest, and the crucial role played by various 
anharmonic resonances at a given energy of interest.  A theoretical approach that incorporates most of these essential findings is the local random matrix theory (LRMT)\cite{schofieldwolynes1994,leitnerwolynes1997,leitnerwolynes98} wherein IVR is 
viewed as diffusion among sites in the zeroth-order quantum number space (QNS). The LRMT model itself arose from earlier studies which mapped the problem of IVR to the phenomenon of Anderson localization\cite{loganwolynes1990}. 
Consequently, and in analogy to the theory of Anderson localization, a scaling approach\cite{schofieldwolynes1993} was developed with specific predictions for the various observables that are relevant for the phenomenon 
of IVR. A key success of the LRMT model has been in terms of providing a criterion for the quantum ergodicity threshold\cite{leitnerwolynes1997} (QET) - below this threshold one has restricted IVR and being above the threshold signals 
facile IVR and, potentially, statistical behaviour. We refer the reader to a recent review\cite{Leitner2015} that highlights the successes of this approach to a number of molecular systems. 

An attractive feature of the LRMT approach to IVR has to do with a close correspondence to the classical phase space descriptions of IVR dynamics. Indeed, recent studies have shown that many of the predictions of the scaling 
approach to IVR hold even for small molecules\cite{semparithi06,manikandan14} at high levels of excitations and with a considerable degree of classical-quantum correspondence. One possible reason for the observation is that the local nature of LRMT 
encapsulates the influence of the various local structures in the classical phase space that modulate the IVR dynamics. It is therefore interesting to ask if QET has a classical counterpart. 
Admittedly, a classical counterpart to QET will not be able to account for dynamical tunneling\cite{hellerdavis81} and other purely quantum effects. Nevertheless, one would have a classical baseline to determine the extent to which pure quantum IVR 
pathways dominate the energy flow out of an initial state. In turn, such an analysis is expected to provide guidelines for the choice of external fields that can alter or control the IVR dynamics. 

In order to address the above question it is essential to study the classical dynamics of models with several degrees of freedom (DOF). In particular, since LRMT is formulated ideally for large molecules, it is necessary to explore 
the detailed phase space structure and their influence on the IVR dynamics for systems with DOF $\geq 3$. Note that there is a rich legacy of classical phase space approach to understanding IVR dynamics in systems with two degrees 
of freedom\cite{uzer91,ezra1992,ezra1998,reinhardt82,Farantos2009,davis1995,toda2005,sibert1jcp82}. These studies have led to considerable insights into IVR in terms of the origins of nonstatistical dynamics due to the various 
dynamical barriers\cite{uzer91,ezra1992,shirts82,davis1985} in phase space, dynamical assignment of highly excited eigenstates\cite{manikandan2009}, and identification of appropriate reaction coordinates in terms of new modes that originate from bifurcations\cite{Kellman2007,Farantos2009,ishikawa1999} in the phase space. However, there are far fewer studies\cite{semparithi06,manikandan14,martens87,Fair1995,shojiguchi07,paskauskas09,ezra2009} that tackle genuinely three or more 
degrees of freedom systems. The main reason for this is that classical conservative systems with DOF $<3$ are not sufficiently general in terms of the phase space topology and transport\cite{LLbook}. For instance, classical phase space structures 
like cantori and the mechanism of stickiness\ in DOF $=2$ cannot be generalized to higher DOF in a straightforward manner\cite{Meiss2015,Bunimovich2008,lange16}. Moreover, the notion of isolated regular regions 
interspersed with chaos is no longer tenable. Instead, the various nonlinear resonances, analogs of the quantum anharmonic resonances, at a given energy of interest form a connected network called as the Arnold web which leads to 
new transport pathways in the multidimensional phase space\cite{wigbook}.  It is only recently that  studies have started to focus on identifying, and visualizing,  dynamical structures\cite{shchekinova04,richter2014,guillery2017,firmbach18} in the multidimensional phase space and 
features on the Arnold web that are important\cite{paskauskas09,lange16,sethi2012,pankaj2015,atkins1992,lopez2016} in regulating the classical IVR dynamics and their fingerprints on the quantum IVR pathways\cite{manikandan14}. 

In this work we explore the IVR dynamics from a classical-quantum correspondence perspective in a $3$-DOF model Hamiltonian inspired by an earlier study on the same system by Martens\cite{ccmjstat92}. Our main focus is to bring out 
the similarity and differences between the classical and quantum IVR pathways on the Arnold web in the vicinity of the so called resonance junctions (cf. section~\ref{cl_web} for details). There are reasons to believe that these 
junctions are the seeds for classically nonstatistical dynamics\cite{manikandan14,pankaj2015,engel1989}. It is therefore of some interest to investigate as to whether the signatures of such resonance junctions manifest in the corresponding quantum dynamics. 
More importantly, in the vicinity of the junctions, quantum dynamics has access to multiple IVR pathways due to dynamical tunneling\cite{kspre05,ksirpc07} - pathways that are not possible in the classical IVR dynamics. 
In fact, various studies\cite{dyntunbook} have implicated both the classical nonlinear resonances, in terms of resonance assisted tunneling (RAT)\cite{almeida84,brodier01,ksjcp05,kspre05,lock2010} and the chaos, in terms of chaos assisted 
tunneling (CAT)\cite{tomsovic94,bohigas93,eltschka05,mouchet06}, towards enhancing dynamical tunneling by several orders of magnitude. At a resonance junction one has 
both nonlinear resonances (in fact, an infinity of them) and chaos. Consequently, it is expected that the  competition between classical and quantum IVR pathways will be subtle and, at the least, the classical-quantum correspondence 
will be tested severely near the resonance junctions. In this context it is relevant to note that a recent study\cite{pittmanheller16} explored the competition  between dynamical tunneling and fast classical diffusion along resonance zones in   a certain class of Hamiltonians  with DOF $\geq 3$.

We begin  with a brief description of the model Hamiltonian and the choice of parameters. Next, we describe our approach to map the Arnold web followed by an overview of the key structures on the web that are of interest to the 
present study. Subsequently, we present a detailed study of the  dynamics near two different resonance junctions and highlight the role of dynamical tunneling in the competition between classical and quantum IVR pathways. We conclude 
with a preliminary study of the nature of the eigenstates near the resonance junctions.

\section{Model system and theoretical preliminaries} \label{modelham}
\label{modham}

The classical $3$-DOF Hamiltonian of interest comes from an earlier study by Martens\cite{ccmjstat92} and has the form
\begin{equation} \label{classham}
 H_{CM}({\bf{J}},{\bm{\theta}}) = H_0({\bf{J}}) + V(\bf{J},{\bm \theta})
\end{equation}
where the action variables ${\bf{J}} = (J_1, J_2, J_3)$ and their conjugate angle variables ${\bm{\theta}} = (\theta_1, \theta_2, \theta_3)$ are associated with $H_{0}$.
The zeroth-order Hamiltonian\cite{ccmjstat92} is given by
\begin{equation} 
H_0({\bf{J}}) = \sum_{i=1}^{3} \bigg[\omega_i J_i + \frac{1}{2} \alpha_i J_i^2 \bigg]
\label{hzero_cm}
\end{equation}
and the perturbation term\cite{ccmjstat92} is chosen as
\begin{equation} 
\begin{aligned}
V(\bf{J},{\bm \theta)} &= 2\beta_1 \sqrt{J^2_1 J_2} \hspace{1mm} \cos(2\theta_1 - \theta_2) \\
                    & + 2\beta_2 \sqrt{J_1^3 J_2^2} \hspace{1mm} \cos(3\theta_1 - 2\theta_2) \\
                    & + 2\beta_3 \sqrt{J_2 J_3^2} \hspace{1mm} \cos(\theta_2 - 2\theta_3)
\end{aligned}
\label{hpert_cm}
\end{equation}
The perturbation term includes three independent nonlinear resonances. The first is a $2$:$1$ resonance (R$_1$), denoted as $(2,-1,0)$, between the first and the second modes, the second term is a $(3,-2,0)$ resonance 
(R$_2$) between the first and the second modes and the third term is a $(0,1,-2)$ resonance (R$_3$) between the second and the third modes. The $\omega_i$'s are the harmonic frequencies, the $\alpha_i$'s are the 
anharmonicities of the model system, and the $\beta_i$'s determine the strength of the nonlinear resonances. The order of any resonance ($k_1, k_2, k_3$) is defined as $\|k_1\| + 
\|k_2\| + \|k_3\|$. Thus, the order of the three resonances R$_{1}$, R$_{2}$, and R$_{3}$ are three, five and three respectively.

The quantum limit of the Hamiltonian eq \ref{classham} is obtained by using the Heisenberg correspondence 
\begin{equation}
 \hat{a} \longleftrightarrow \sqrt J \hspace{1mm} e^{i\theta}\,\,,\,\, \hat{a}^{\dagger} \longleftrightarrow \sqrt J \hspace{1mm} e^{-i\theta}
\end{equation}
where $\hat{a}$, $\hat{a}^{\dagger}$ are the harmonic annihilation and creation operator respectively. Thus, the 
corresponding quantum Hamiltonian is 
\begin{equation} 
 \hat{H}_{QM} = \hat{H}_0 + \hat{V}
 \label{qumham}
\end{equation}
with the zeroth-order term being
\begin{equation} \label{qumh0}
\hat{H}_0 = \sum_{i=1}^{3} \bigg[\omega_i \Big(\hat{a}_i^{\dagger}\hat{a}_i + \frac{1}{2}\Big) + \frac{1}{2} \alpha_i \Big(\hat{a}_i^{\dagger}\hat{a}_i + \frac{1}{2}\Big)^2 \bigg]
\end{equation}
and the resonant coupling given by
\begin{equation} \label{qumv}
\begin{aligned}
 \hat{V} & = \beta_1 [(\hat{a}_1^{\dagger})^{2}  \hat{a}_2 + \hat{a}^{\dagger}_2 (\hat{a}_1)^{2}  ] \\
      & + \beta_2 [(\hat{a}_1^{\dagger})^3 (\hat{a}_2)^2 + (\hat{a}_2^{\dagger})^2 (\hat{a}_1)^3] \\
      & + \beta_3 [\hat{a}_2^{\dagger} (\hat{a}_3)^2 + (\hat{a}_3^{\dagger})^2 \hat{a}_2]
\end{aligned}
\end{equation}
 Note that the above Hamiltonian is an example of an effective spectroscopic Hamiltonian with $H_{0}$ being the Dunham expansion and $V$ representing the dominant Fermi resonance terms in the energy range of interest. Although 
 there are certain drawbacks, when compared to analyzing the dynamics on exact potential energy surfaces, such Hamiltonians have been widely utilized to understand the spectroscopy and dynamics in the highly excited energy regimes. 
 As an aside we mention that Hamiltonians of the above form also arise in the context of trapped Bose-Einstein condensates\cite{spekkens99,khan03, khan05} and, from a nonlinear dynamics point of view, can be thought of as resonant normal forms.  

\subsection{The Arnold web: definition and construction using fast Lyapunov indicator} 
\label{cl_web}

For a general classical Hamiltonian
\begin{equation}
 H = H_0({\bf{J}}) + \lambda V(\bf{J},{\bm \theta})
\end{equation}
the nonlinear frequency of the $i^{\rm th}$ mode is defined as 
\begin{equation} \label{nonlinfreq}
\begin{aligned}
 \Omega_i &= \dfrac{\partial H_0({\bf{J}})}{\partial J_i} + \lambda \dfrac{\partial V(\bf{J}, {\bm \theta})}{\partial J_i} \\
          &= \Omega^{(0)}_i ({\bf{J}}) + \lambda \dfrac{\partial V}{\partial J_i}
\end{aligned}
\end{equation}
For an integrable system ($\lambda = 0$), the zeroth-order nonlinear frequencies $\Omega_i \equiv \Omega_i^{(0)}$ are constant since the actions are constants of the motion. However, for $\lambda \neq 0$, $\Omega_i$ the system is non-integrable 
and the various $\Omega_i$ vary with time since the actions are no longer conserved. For the near-integrable case the condition 
\begin{equation} 
 {\bf k}^{r} \cdot {\bm \Omega}^{(0)} = {\bm {0}}
 \label{res_cond}
\end{equation}
with ${\bf k}^r = (k^{r}_{1},k_{2}^{r},k_{3}^{r},\ldots)$ being an integer vector with coprime entries, corresponds to the various resonances $r = 1, 2, \ldots$ that can manifest on the energy shell $H_0 = E$ of interest. Given the zeroth-order energy surface, the above condition for resonances define a 
hypersurface in the ${\bf{J}}$-space. Each resonance hypersurface may intersect with $H_0({\bf{J}}) = E$ and the collection
\begin{equation} \label{res_set}
 \{ (H_0 = E) \cap ({\bf{ k}^{r} \cdot {\bm \Omega^{(0)}}} = {\bf{0}}) \}
\end{equation}
forms a dense web, known as the Arnold web. Particularly, for the model system of interest, as the $H_0({\bf{J}})$ is quadratic in actions, the resonance 
condition defined in eq \ref{res_cond} is a plane and eq \ref{res_set} represents a collection of lines.

The above picture is valid in the near-integrable regime. In general one may still look at the collection 
\begin{equation}
 \{ (H = E) \cap ({\bf{k}}^{r} \cdot {\bm{\Omega}} = {\bm{0}}) \}
\end{equation}
with ${\bm \Omega}$ being an appropriate set of frequencies. Note that ${\bm \Omega}$ have to be determined numerically and can be strictly defined only in the case of regular dynamics. In case of mixed regular-chaotic dynamics the ${\bm \Omega}$ are time dependent and there are subtleties associated with extracting the appropriate frequencies and their interpretations. Nevertheless, time-frequency analysis\cite{martens87,arevalo01,chanwigguzer03} of even the irregular trajectories contain valuable dynamical information. In the present study we do not perform any time-frequency analysis but map the Arnold web using a different, but equivalent, approach. In any case, determining the relevant resonance network is crucial since it provides a visualization of the global phase space structure and the classical IVR dynamics is entirely characterized by such a network.

 There are several ways to map the Arnold web for a system, including the frequency modulation indicator\cite{cordani2008} 
and methods based on the various chaos indicators\cite{froeschlelega97, froeschlelega00, barriospu09, cincotta11}. In this work we use the method based on the fast Lyapunov indicator\cite{froeschlelega97, froeschlelega00} 
(FLI) which is  sensitive to the various 
phase space structures and can distinguish between chaotic and regular trajectories. Moreover, among the regular trajectories, FLI can pick out a resonant trajectory from a non-resonant one. 
Consider a dynamical system defined in general by
\begin{equation} \label{dyn_sys}
    \frac{d\bf{X}}{dt} = \bf{F}(\bf{X})
\end{equation}
where ${\bf{X}}=(x_1,x_2,...,x_n)$ is the set of dynamical variables. The FLI is then computed by solving the variational equations,
\begin{equation} \label{var_eqn}
    \frac{d\bf{v}}{dt} = \Big(\frac{\partial \bf{F}}{\partial \bf{X}} \Big) \bf{v}
\end{equation}
where $\bf{v}$ is any n-dimensional vector (n is the dimension of the phase space). The FLI defined as
\begin{equation} \label{fli}
    \text{FLI}({\bf{X}}(0),{\bf{v}}(0),t) = \log||{\bf v} (t)||
\end{equation}
is distinctively different for chaotic and regular trajectories provided the time is sufficiently long.
However, the variation of FLI with time show rapid oscillations and hence 
the distinction between resonant and non-resonant regular trajectories becomes difficult. To avoid this another definition of FLI was introduced by Lega and Froeschl\'e \cite{legafroeschle01} as

\begin{equation} \label{mod_fli}
    \text{FLI}({\bf{X}}(0),{\bf{v}}(0),T) =  \sup_{0<k<T} \log||{\bf v} (t)||
\end{equation}
In other words, at any time $t$, the modified FLI eq \ref{mod_fli} is defined as the maximum value of FLI in the time interval $(0,t)$. We mention here that an arbitrary choice for the vector ${\bf v}$ in eq~\ref{fli} can result in the appearance of spurious structures on the web\cite{barriospu09}. Thus, in the present work we choose this vector according to the prescription given by Barrio et al\cite{barriospu09}.

\subsection{Three dynamical regimes} 
\label{dynregimes}

In the classical dynamical description of systems, particularly with DOF $ \geq 3$, one usually associates three dynamical regimes of interest --  Kolmogorov-Arnold-Moser (KAM), Nekhoroshev, and Chirikov.  
In order to give a brief description of the different regimes, consider a Hamiltonian $H({\mathbf J},{\bm \theta}) = H_{0}({\mathbf J}) + \lambda V({\mathbf J},{\bm \theta})$ with $H_0$ being the unperturbed Hamiltonian. In the KAM regime almost all the unperturbed tori remain intact with small deformations. The measure of destroyed tori, corresponding to resonant tori, is negligible. In other words, for a finite perturbation with $\lambda \ll 1$ it is possible to transform the Hamiltonian to an effective Hamiltonian which is integrable upto fairly high orders in the perturbation strength $\lambda$. In the Nekhoroshev regime, the measure of destroyed resonant tori is not negligible, but still small. In this case, the resonances form a connected network called as the Arnold web. However, one can still distinguish the regions with unperturbed invariant tori from the resonant regions formed by the tori that are destroyed. With further increase in $\lambda$ and for some $\lambda > \lambda_0$ the resonance regions grow, overlap and create broad regions of chaos. There is no place for unperturbed invariant tori and large parts of the Arnold web lose their structure. This regime signals the onset of the Chirikov regime. 

Note that the above description of the different regimes is not rigorous. A more formal treatment\cite{harsoula2013} would require identifying the total order to which resonances need to be considered and the resulting approximate partitioning of the Arnold web into domains with no, single, and multiple resonances. The highest order of resonance considered also translates into an appropriate largest timescale of interest. Since our model system eq~\ref{hpert_cm} is already  a resonant normal form, one can think that such a choice of the key resonances has been already made. For our present purpose it is sufficient to associate the different regimes with the extent of classical energy flow,  ranging from a highly restricted IVR to fairly facile IVR. The nature of the phase space transport, in other words classical IVR dynamics, is significantly different in these regimes. Whereas in the KAM regime there is very little transport, in the Nekhoroshev regime it is possible for transport to occur over extremely long (exponential), often physically irrelevant, timescales. In contrast, the Chirikov regime signals facile transport over large regions of the phase space facilitated by the various resonances that have overlapped.  

It seems reasonable to associate the passage through the three dynamical regimes, as the perturbation strength $\lambda$ is tuned, with the classical analog of the onset of QET. Indeed, the key ingredients of QET like the effective resonant strengths and the state space connectivity do have a classical interpretation. However, a firm answer to whether there is a classical analog of QET requires a careful study of the competition between classical and quantum IVR pathways in the various dynamical regimes. The results presented in section~\ref{results} is our first attempt to address this important issue. 

\section{Results and discussions}
\label{results}

We start with our choice for the relevant parameters of the model system and the computational approach adopted to study the classical and quantum IVR dynamics. The specific choice of parameters are motivated in terms of taking the system from the  KAM through Nekhoroshev to the Chirikov regime. The parameter values considered here should be taken as a representative example for a class of Hamiltonians whose dynamics  can be essentially determined in terms of three independent resonances which intersect to form junctions on the Arnold web.

\subsection{Parameters of the model and computational details}

The parameters for the model Hamiltonian eq \ref{classham} are chosen in such a way that the primary nonlinear resonances intersect with each other at the energy of interest and form resonance junctions. 
We take $(\omega_1, \omega_2, \omega_3) = (1.1, 1.7, 0.9)$ and $(\alpha_1,\alpha_2,\alpha_3) = (-0.0125,  -0.02,  -0.0085)$ as parameter values for the zeroth order Hamiltonian. The strength of the nonlinear resonances can be tuned by changing the 
$\beta_i$ parameters.  The different sets of the resonant coupling strengths will be represented as $[\beta_1,\beta_2,\beta_3] \times \epsilon$, 
with a fixed $\epsilon = 10^{-5}$.  For convenience, in the rest of the paper the coupling parameters will be denoted as [$\beta_1$, $\beta_2$, $\beta_3$], suppressing the $\epsilon$ factor.  For illustrating our main findings, we focus on a total energy $E = 40$. Note that the dissociation energies of the three Morse oscillators in eq~\ref{hzero_cm}
are $(D_1,D_2,D_3) = (48.4, 78.2, 47.6)$ for the given choice of the harmonic frequencies and the anharmonicities. Thus, with $E=40$ we are studying the IVR dynamics at high excitation energies which, in molecular systems, 
is typical of IVR near the dissociation thresholds and hence of particular relevance to reaction dynamics.

To analyze the classical dynamics, the Hamilton's equations of motion are integrated using an $8^{\rm th}$ order Runge-Kutta method. 
For the quantum mechanical studies, we use the zeroth-order number basis $\{|{\bf{n}} \rangle \} \equiv \{|n_1, n_2, n_3 \rangle\}$ to diagonalize the quantum Hamiltonian $\hat{H}_{QM}$ to obtain the eigenvalues and the 
corresponding eigenstates. Since there are no conserved polyads for the system, thus preventing an efficient block-diagonalization procedure, the size of the Hamiltonian matrix can become quite unwieldy. 
However, since our interest is in studying the IVR dynamics and eigenstates in a specific energy range of interest, we adopt the following strategy. To converge the eigenstates  in the range $E \in  (39.5, 40.5)$, 
a sufficiently large number basis with zeroth-order energies centred at $E = 40$ is taken and the 
number of basis states are increased until the eigenvalues are converged up to seventh decimal place. Care is taken such that a sufficiently large number of the eigenstates are converged to faithfully capture the 
IVR dynamics in the energy range and timescales of interest. All the quantum calculations reported here are for $\hbar$ =1.

In this work we use eq \ref{mod_fli} to compute the Arnold web in the action space at a total energy of E = 40. To construct the web, a suitably large grid of initial actions in $(J^{(0)}_1,J^{(0)}_2)$ space is chosen along with a fixed set of initial angles $(\theta^{(0)}_1,\theta^{(0)}_2,\theta^{(0)}_3)$. The value of the  remaining initial action $J^{(0)}_{3}$ is
then obtained, if allowed, from the constraint of energy conservation $H(J^{(0)}_{3}; J^{(0)}_1,J^{(0)}_2,{\bm \theta}^{(0)}) = E$. Trajectories with initial conditions $({\bf J}^{(0)},{\bm \theta}^{(0)})$ are then integrated for a total time of $T = 5000$ units, corresponding to about $715$ harmonic periods of the lowest frequency mode, and the corresponding FLI are computed.

For the Arnold webs shown in this work we have chosen the angle slice $(\theta^{(0)}_1,\theta^{(0)}_2,\theta^{(0)}_3) = (\pi/2,\pi/2,\pi/2)$. Note that there is nothing special about the chosen angle slice and other slices can be used as well. 
In the  near-integrable regimes the Arnold web structures persist in any arbitrary choice of angle slice. However, far from near-integrability,  different angle slices of the phase space can reveal different structures and one does not a priori know whether there is an optimal choice.  The present work does not attempt to explore this, no doubt important, aspect further since the classical and quantum results are both compared on the  chosen action space projection. 

\begin{figure}[!t]
\begin{center}
\includegraphics[width=1\linewidth]{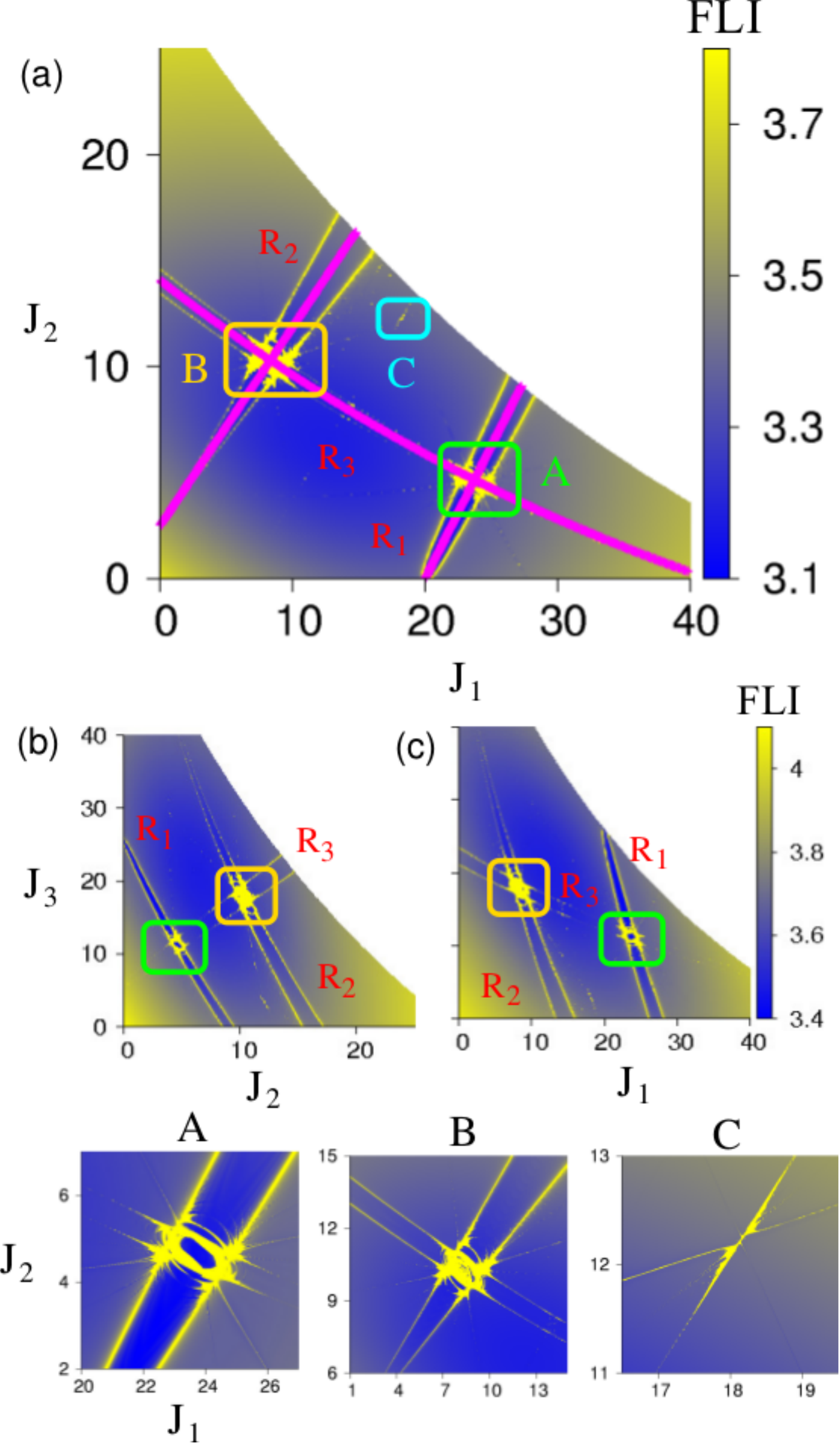}
\caption{(a) The Arnold web for resonant coupling strengths $[\beta_1,\beta_2,\beta_3] = [5,1,5]$ projected on to the  $(J_1,J_2)$ space. The three primary resonances in Hamiltonain eq~\ref{classham} are denoted by R$_1$, R$_2$ and R$_3$.
The zeroth-order resonance lines determined using eq~\ref{res_cond} are shown in magenta color. The green (A) and gold (B) color boxes 
are the two primary resonance junctions and the cyan (C) box is another resonance junction formed from induced resonances. (b) and (c)  Different state (action) space projections of the Arnold web in (a). The bottom panel shows a close up of the regions near the resonance junctions A, B, and C. Each web is constructed by choosing a $500 \times 500$ grid of initial conditions. See text for details.}
\label{fig:web_5_1_5}
\end{center}
\end{figure}

\subsection{Overview of the Arnold web structures} 
\label{cl_web_structure}

In the  absence of couplings i.e., $\beta_i = 0$ in eq \ref{hpert_cm}, the system is integrable and the phase space is filled with invariant tori. For sufficiently small couplings, 
according to the KAM theorem, the tori with irrational frequency ratio remain invariant. The tori with rational frequency ratios are destroyed and form resonance zones with widths depending on the order of the resonance.  At the crossings of 
two or more independent resonance zones, junctions are formed. Note that the junction formed by the intersection of $k$ independent resonances is called as multiplicity-$k$ junction. In three 
DOF one can have only multiplicity-$2$ junctions. Hence, all the junctions studied in this work are multiplicity-$2$ junctions.

In Figure \ref{fig:web_5_1_5}(a), the Arnold web for coupling $[5,1,5]$ is shown in the $(J_1,J_2)$ plane. The zeroth-order resonance lines, predicted using the nonlinear zeroth-order frequencies 
eq \ref{nonlinfreq} are also shown for comparison. For reference, in Figure \ref{fig:web_5_1_5}(b) and (c)  the same web is shown in the $(J_2,J_3)$ and the $(J_1,J_3)$ spaces respectively. On inspecting Figure \ref{fig:web_5_1_5}(a) it is clear that two prominent resonance junctions appear at the energy of interest. One of the junctions, labeled A,  is formed from the intersection of the $(2,-1,0)$ and the  $(0,1,-2)$ resonance zones wheres the other junction, labeled B, involves the crossing of   the $(3,-2,0)$ and $(0,1,-2)$ resonance zones. In the rest of the paper these junctions will be referred to as junction-A and junction-B. The presence of yellow regions near the junctions indicate the appearance of chaos.  There is yet another resonance junction which is labeled as C in Figure \ref{fig:web_5_1_5}(a). This particular junction is formed from the intersection of two induced resonances emanating from the junctions A and B.  A close up view of the junctions and the finer details can be seen in  Figure \ref{fig:web_5_1_5}A-C which show the presence of some of the induced resonances near the junctions. In this work we will focus on junctions A and B only.

\begin{figure}[t]
\begin{center}
\includegraphics[width=\linewidth]{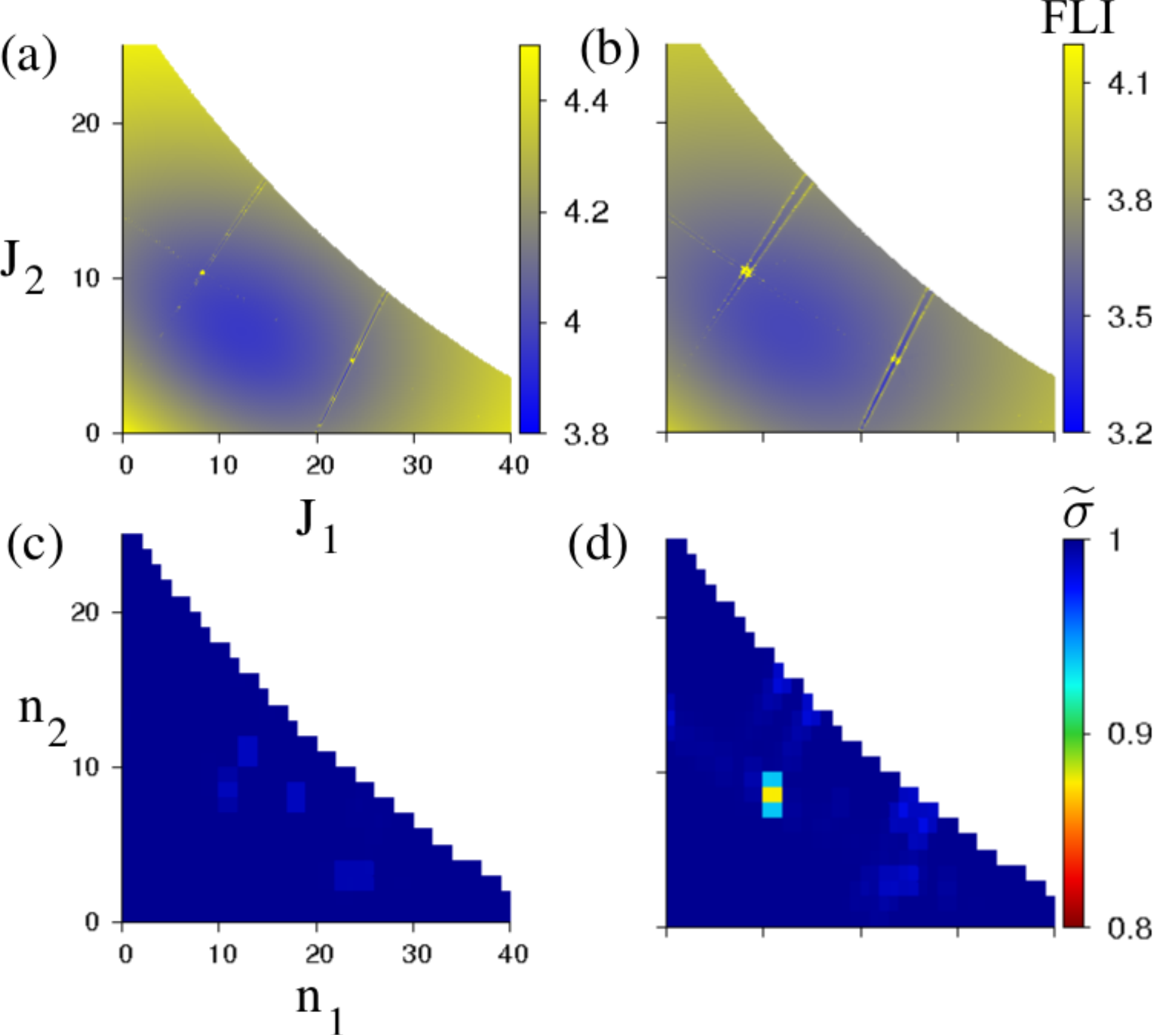}
\caption{The Arnold web at $E=40$ for two different coupling strengths  $[\beta_1,\beta_2,\beta_3]$. (a) $[0.1,0.01,0.1]$ and (b) $[0.5,0.1,0.5]$. The corresponding dilution factor (eq.~\ref{redsigma}) plots in the quantum number space are shown in (c) and (d) respectively.}
\label{fig:kam_nekh}
\end{center}
\end{figure}

As mentioned in section~\ref{dynregimes}, there are three different regimes of the $3$-DOF phase space that appear with increasing 
couplings. 
In Figure \ref{fig:kam_nekh}(a)-(b), as examples, the Arnold web for small coupling values  $[0.1,0.01,0.1]$ and $[0.5,0.1,0.5]$ corresponding to the KAM and the Nekhoroshev regimes 
are shown. In the KAM regime shown in Figure \ref{fig:kam_nekh}(a) two of the resonances can be seen and the rest of the phase space is filled with KAM 
tori. In Figure \ref{fig:kam_nekh}(b) chaos is clearly observed at the resonance junctions, but the chaotic regions near the two junctions are not connected, and hence this case corresponds to the Nekhoroshev regime. Note that our classification of the webs in Figure~\ref{fig:kam_nekh} is by no means accurate and is meant only as a rough guide to these two regimes. In particular, although there are other interesting dynamical issues to be addressed, the IVR  in these regimes is highly restricted and hence will not be of interest to us in the present work. We therefore focus on slightly larger coupling strengths for the rest of the study.

In  Figure \ref{fig:web_qumweb}(a) we show the Arnold webs at $E=40$ for increasing values of the resonant coupling strengths and one can observe that
 the widths of the resonances increase and considerable chaos starts to appear at 
the junctions. Note that for the coupling strength of $[5,1,5]$, shown  in Figure \ref{fig:web_qumweb}(a), the system already shows nearly connected chaotic layers and this connected chaotic layer becomes quite prominent for coupling strength $[20,1,20]$ wherein  the system is in the Chirikov regime. For even 
larger values of the coupling, e.g. $[50,5,50]$ and $[200,10,200]$, the different resonances  overlap extensively with each other leading to wide spread chaos and hence into the deep 
Chirikov regime.

\subsection{Does the quantum state space sense the Arnold web?} 
\label{qum_web}

\begin{figure}[t]
\begin{center}
\includegraphics[width=0.5\textwidth]{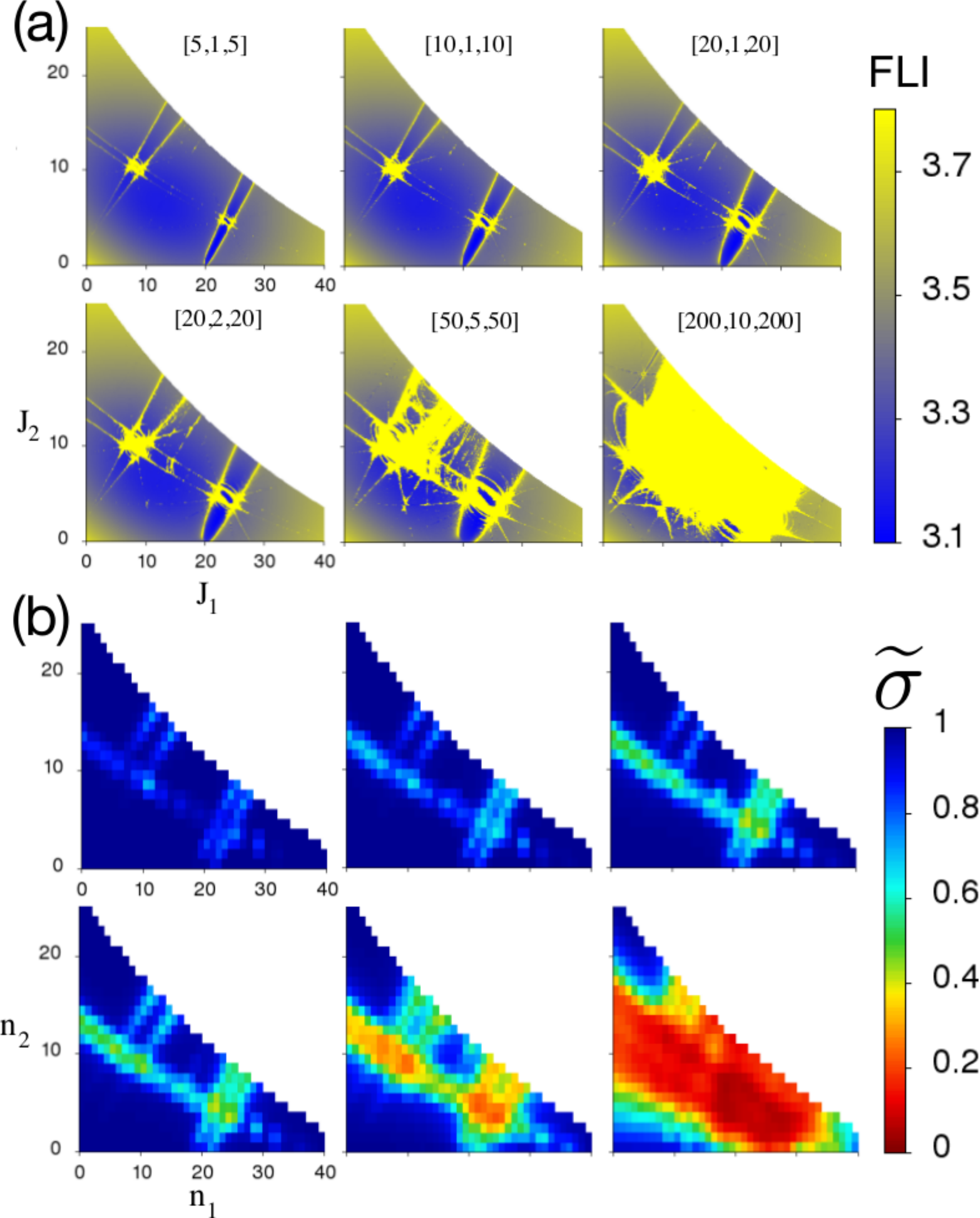}
\caption{(a) The Arnold web for different values of $[\beta_1,\beta_2,\beta_3]$ (indicated) at energy E = 40. (b) The corresponding dilution factor plots (eq.~\ref{redsigma}). See text for discussions and details.}
\label{fig:web_qumweb}
\end{center}
\end{figure}

Classically, the dynamics is visualized using the Arnold web in the zeroth-order action space. From the semiclassical quantization rule, it is well known that the classical action $(J)$ and the quantum number $(n)$ are related by the relation, $ J \leftrightarrow (n + 1/2) \hbar$. Thus, the quantum analog of the classical zeroth-order action space is the quantum number space (QNS) or the quantum state space\cite{mgrub00,mgrubpwol04}. As mentioned in the introduction, since LRMT and the scaling theory view IVR as a diffusion in the QNS, the classical-quantum correspondence between actions and quantum numbers suggest that the structures observed in the Arnold webs in Figure~\ref{fig:kam_nekh} and Figure~\ref{fig:web_qumweb} should manifest in the QNS as well. More so, since LRMT is based on the local coupling of an initial quantum zeroth-order state (ZOS) to other states due to the anharmonic resonances. In this section we provide one possible way to realize such a correspondence. 

A widely used quantum measure for the extent of mixing of a certain zeroth-order state  due to IVR is the dilution factor\cite{macdonald83,sibertmgrub04}. 
The dilution factor of a given ZOS $|{\bf{n}} \rangle = \sum_i c_{i {\bf n}} |\psi_i \rangle $, with $c_{i {\bf n}} \equiv \langle {\bf{n}}|\psi_i \rangle$ and $ |\psi_i \rangle $ being the eigenstates of eq~\ref{qumham}, is defined as
 
\begin{equation} \label{dil_fac}
 \sigma_{\bf{n}} = \sum_i |\langle {\bf{n}}|\psi_i \rangle|^4 \equiv \sum_i |c_{i {\bf n}}|^{4}
\end{equation}
Smaller the value of $\sigma_{\bf{n}}$ more coupled the ZOS is to the other ZOS. Since the $\sigma_{\bf{n}}$ is a measure of the mixing due to the anharmonic resonances, one may surmise that they might encode the
information about the Arnold web. In order to confirm this, we 
project the $\sigma_{\bf{n}}$ on to the $(n_1,n_2)$ space as follows. For a given $n_1$ and $n_2$, all possible $n_3$ values with zeroth-order energy $E_0$ in the energy range $(39.5,40.5)$  of interest are considered and an average value of $\sigma_{\bf{n}}$ is calculated

\begin{equation}
 \widetilde{\sigma}(n_1,n_2) = \frac{1}{N} \sum_{n_3} \sigma_{\bf{n}}
 \label{redsigma}
\end{equation}
where $N$ is the total number of allowed $n_3$ subject to the energy criterion. The resulting $\widetilde{\sigma}(n_1,n_2)$ is then associated with a point in the projected QNS.

In Figure \ref{fig:kam_nekh}(c)-(d) we show the  $\widetilde{\sigma}$  when the system is in the KAM and the Nekhoroshev regime. In both the 
cases, the $\widetilde{\sigma}$ does not feel the presence of the phase space structures. 
Figure \ref{fig:web_qumweb}(b) show the results for the couplings corresponding to those of Figure \ref{fig:web_qumweb}(a) and it is clear that  the  
$\widetilde{\sigma}$ representation of the QNS bears close resemblance to the classical Arnold webs.  In these coupling regimes, like the FLI for classical Arnold web, the  
$\widetilde{\sigma}$  can distinguish between different ZOS in terms of the extent of 
their local coupling due to the anharmonic resonances. Thus, the measure $\widetilde{\sigma}$ can be loosely associated with a ``quantum web'' at the energy of interest. 
In the KAM region the ZOS are close to being  quantum 
eigenstates even for the perturbed Hamiltonian with $\widetilde{\sigma} \sim 1$ reflecting localized eigenstates. Near a single resonance, due to 
resonance assisted tunneling\cite{brodier01}, different ZOS can get coupled 
and typically $\widetilde{\sigma} < 1$. Near  resonance junctions it is expected that the multiple resonances with their quantum 
manifestations would lead to $\widetilde{\sigma} \ll 1$.

Note that at $\hbar$ =1, the $\widetilde{\sigma}$ does indicate the presence of junction-A and 
junction-B and can clearly distinuguish the KAM regions from the resonant regions. However, as in the case of Figure \ref{fig:kam_nekh}(c)-(d), junction-C is not detected. This is related to the quantum coarsening of the sub-Planck structures and we anticipate that 
reducing the effective $\hbar$ would reveal these finer structures. Such a correspondence has also been observed\cite{qumNekho16}, using a different approach, in a recent study  wherein  the quantum manifestation of Nekhoroshev stability has been established. Thus, our results in Figure \ref{fig:web_qumweb}(b) show that the generalization may hold for the higher dimensional systems as well.

\subsection{Classical and quantum IVR dynamics} 
\label{cl_qm_IVR}

We now study the dynamics of specific classes of initial ZOS both classically and quantum mechanically. 
For the classical calculation we choose $1000$ initial random angles for a fixed choice of the initial actions corresponding to the quantum ZOS of interest.  
The classical survival probability of an initial state $(J^{(0)}_1, J^{(0)}_2, J^{(0)}_3)$ at a time $t = t^{'}$ is calculated as the fraction of trajectories with action $(J_1(t^{'}), J_2(t^{'}), J_3(t^{'}))$ such that, $|J_i(t^{'}) - J^{(0)}_i|\le 0.5$ irrespective of the angles. The quantum evolution of  an initial ZOS  
\begin{equation}
 \begin{aligned}
 |{\bf n}(t) \rangle & = e^{-i\hat{H}t/\hbar} |{\bf n}(0) \rangle \\
                    & = \sum_{k} |\psi_{k} \rangle \langle \psi_{k} |e^{-i\hat{H}t/\hbar} |{\bf n}(0) \rangle \\
                    & = \sum_{k} c_{{\bf n} k}  \, e^{-iE_{k}t/\hbar} |\psi_{k} \rangle \\
 \end{aligned}
\end{equation}
yields the survival probability $P_{\bf{n}}(t)$ of the state  as 
\begin{equation} \label{qum_surv_prob}
 \begin{aligned}
  P_{\bf{n}}(t) & \equiv |\langle {\bf{n}}(0)|{\bf{n}}(t)\rangle|^2 \\
                & = \bigg| \sum_{k} |c_{{\bf n} k}|^2  e^{-iE_{k}t/\hbar} \bigg|^2 \\
 \end{aligned}
\end{equation}
where $E_{k}$ and $|\psi_{k} \rangle$  are the eigenvalues and the eigenstates of eq~\ref{qumham} respectively.

\begin{figure}[t]
\begin{center}
\includegraphics[width=1\linewidth]{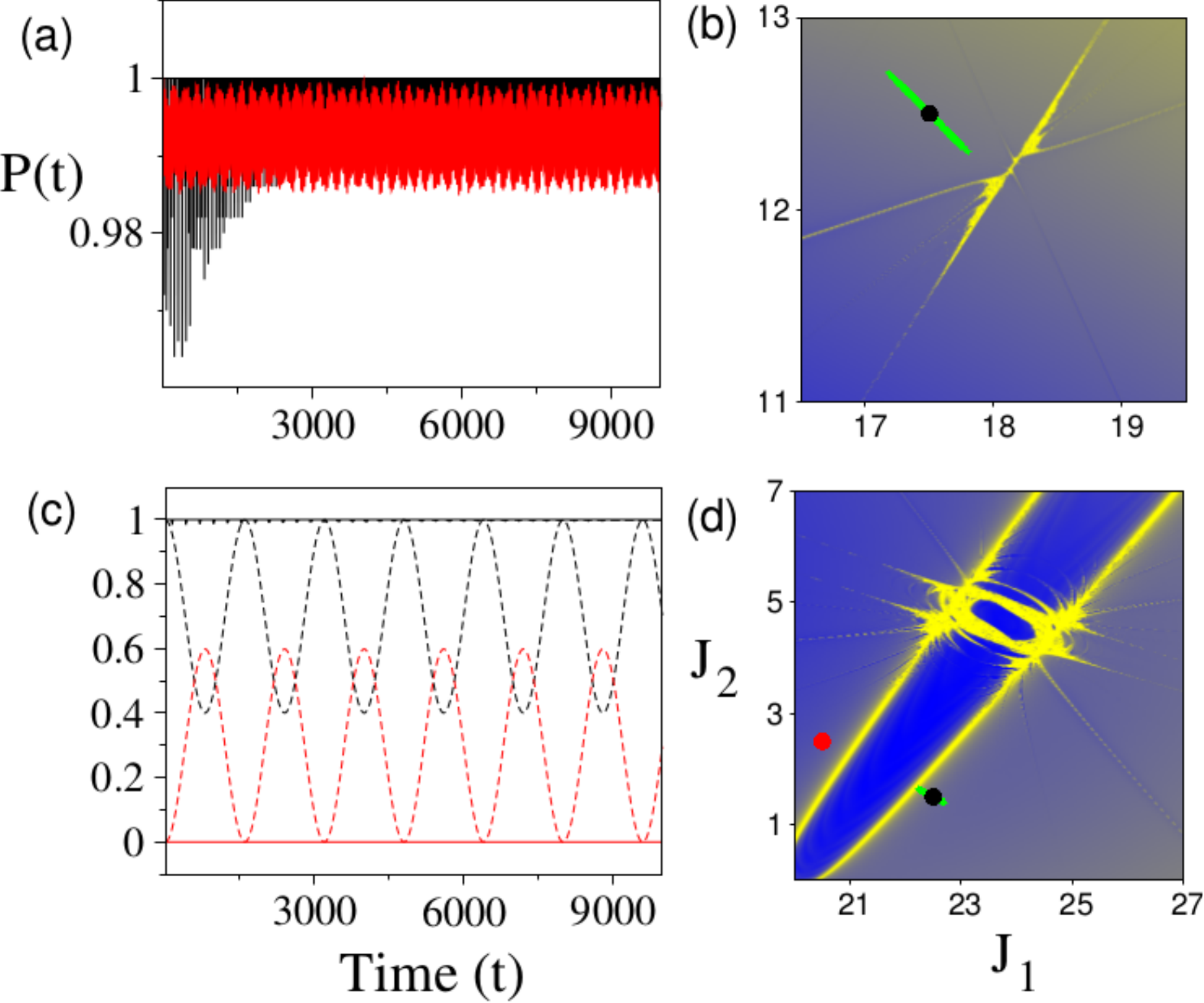}
\caption{(a) Survival probability, classical (black) and quantum (red), of an initial state $|17,12,3\rangle$. (b) Location (black dot) of $|17,12,3\rangle$ on the web. The classical motion projected on the web is shown as green points. 
(c) Survival probability of the state $|22,1,19\rangle$ (black) and cross survival probability of the state $|20,2,19\rangle$ (red). Both classical (solid lines) and quantum (dotted lines) results are shown (d) Location of $|22,1,19\rangle$ (black dot) 
and $|20,2,19\rangle$ (red dot) and the classical motion of $|22,1,19\rangle$ projected on on the web (green points) are shown.}
\label{fig:kam_res_dyn}
\end{center}
\end{figure}

\subsubsection{Dynamics in the KAM region} 
\label{KAM_dyn}

As is clear from the ``quantum web'' shown in Figure \ref{fig:kam_nekh}(c)-(d), the dilution factor of a ZOS located in the KAM regime is nearly one, implying little local coupling with other states.  In Figure \ref{fig:kam_res_dyn}(b), the location of a state $|17,12,3\rangle$ belonging to this class of states is shown and is seen to be located near the junction-C in Figure~ \ref{fig:web_5_1_5}. Also shown in the figure is the 
classical evolution projected on the angle slice of choice. The projected classical evolution, akin to what is done in the surface of section studies in lower DOF cases, on a specific angle slice  $(\theta_1^{(c)}, \theta_2^{(c)}, \theta_3^{(c)})$ is obtained by plotting the actions of 
only those trajectories for which $|\theta_i^{(c)} - \theta_i| \le 0.1$, for all $i = 1, 2, 3$.  As expected, Figure~\ref{fig:kam_res_dyn}(a) shows that the classical survival probability does not decay and the quantum survival probability behaves similarly. Note that $|17,12,3\rangle$ is shown as an example and other states belonging to this class  show similar behaviour. 

\subsubsection{Dynamics near a single resonance} 
\label{sin_res_dyn}

In Figure \ref{fig:kam_res_dyn}(c) we compare the classical and quantum survival probabilities of a ZOS $|22,1,19\rangle$ which is located away from the junction-A 
but close to the $(2,-1,0)$ resonance. Clearly, there is a significant difference since  classically there is very little IVR whereas the quantum probability decays significantly and exhibits a beating pattern. The beating pattern is suggestive of a partial two-state mixing and characteristic of a RAT process\cite{brodier01}. A closer examination reveals that the state $|20,2,19\rangle$ mixes with the initial ZOS and, as evident from Figure \ref{fig:kam_res_dyn}(d), the two states are located on the opposite sides of the $(2,-1,0)$ resonance zone.
The quantum mixing is not an ideal $2$-state case since the states are not symmetrically located about the resonance center.
Note that Figure~\ref{fig:kam_res_dyn}(c) also shows that the classical 
probability of $|22,1,19\rangle$  does not flow into $|20,2,19\rangle$ which is further confirmed by inspecting the projected classical dynamics on the web in Figure~\ref{fig:kam_res_dyn}(d). Hence the key difference between classical and quantum IVR dynamics in this case is due to dynamical tunneling. The results for $|22,1,19\rangle$, and other similar cases involving ZOS near single resonances, shown  here are not surprising since one can effectively reduce the $3$-DOF Hamiltonian to a single degree of freedom pendulum Hamiltonian that accounts for the observed regular dynamics and the RAT phenomenon.

\begin{figure*}[t]
\begin{center}
\includegraphics[width=1\linewidth]{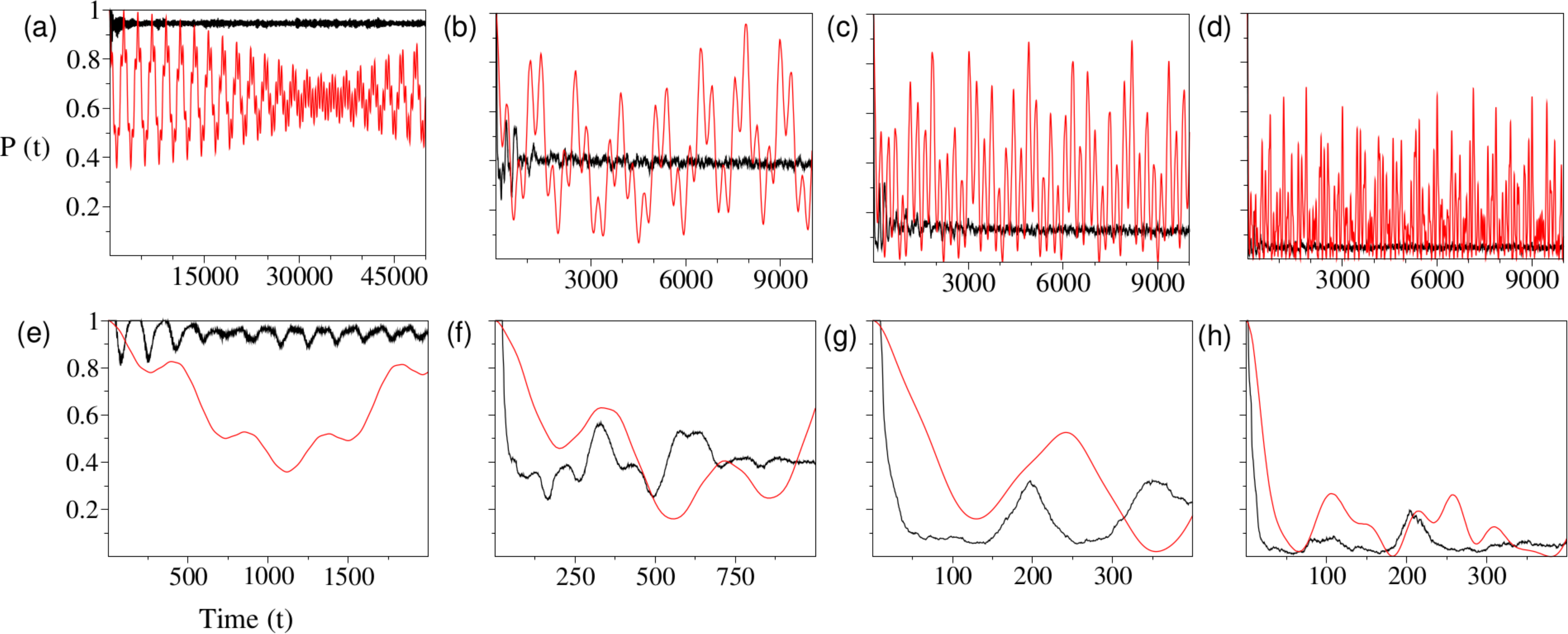}
\caption{Classical (black) and quantum (red) survival probabilities of the initial ZOS $|25,4,9\rangle$ for varying coupling strengths $[\beta_1,\beta_2,\beta_3]$  (a) $[5,1,5]$ 
(b) $[10,1,10]$ (c) $[20,1,20]$ (d) $[50,5,50]$. The bottom panel shows the corresponding plots on short time scales.}
\label{fig:sp_25_4_9}
\end{center}
\end{figure*}

\subsubsection{Dynamics near a resonance junction} 
\label{junc_dyn}

Dynamics of ZOS near the resonance junctions is supposed to be the most complicated as it is coupled to many other states locally. On one hand, near the junction, presence of more than one independent resonance provides several resonance-assisted pathways for dynamical tunneling. On the other hand, increasing resonance coupling strength leads to increased chaos in the vicinity of the junction and increased classical transport. One therefore expects a subtle classical-quantum competition in the IVR dynamics since the chaos will mix certain states classically and the classical chaos can influence the dynamical tunneling process in terms of a combination of RAT and CAT\cite{bohigas93}.

\begin{figure*}[t]
\begin{center}
\includegraphics[width=\textwidth]{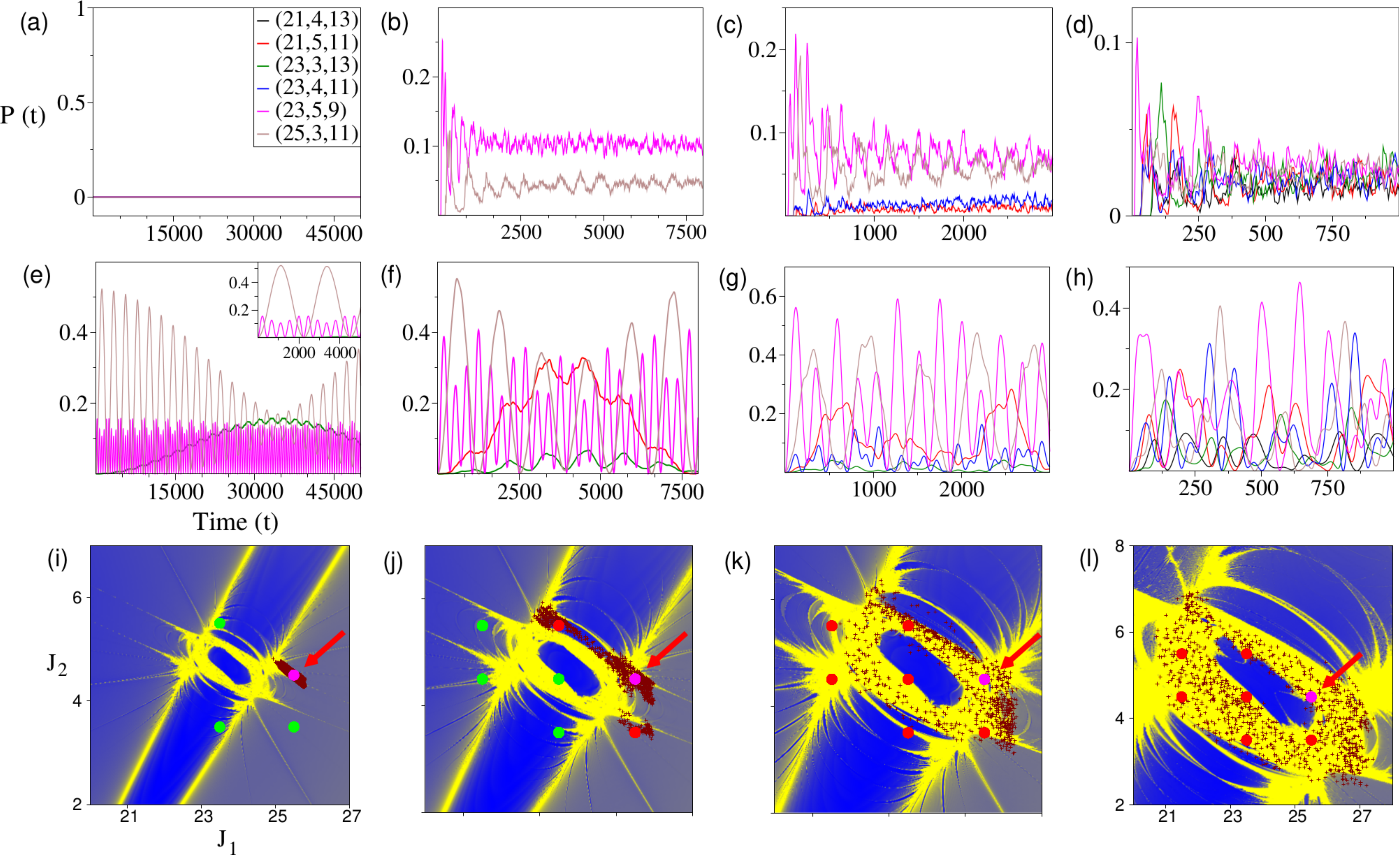}
\caption{Classical cross survival probabilities (top panel) of the initial ZOS $|25,4,9\rangle$ for different coupling strengths $[\beta_1,\beta_2,\beta_3]$ (a) $[5,1,5]$
(b) $[10,1,10]$ (c) $[20,1,20]$  (d) $[50,5,50]$. The corresponding quantum results are shown in the middle panel. The legend in (a) indicates the states of interest. 
The bottom panel shows the initial ZOS (arrow), classical dynamics of the ZOS (maroon points, upto total time $T=10000$), and the location of the mixing states in the vicinity of junction-A on the Arnold web.
Note that states which are mixing only quantum mechanically are shown 
as green dots and the states which mix both classically and quantum mechanically are shown as red dots. See text for details.}
\label{fig:csp_25_4_9}
\end{center}
\end{figure*}

In order to bring out this complexity, in Figure~\ref{fig:sp_25_4_9}(a)-(d) we compare the classical and quantum survival probabilities of an initial ZOS $|25,4,9\rangle$, located near junction-A,  with increasing resonant coupling strengths. 
Results for the smallest coupling value shown in Figure~\ref{fig:sp_25_4_9}(a) indicate that the classical probability does not decay much but the quantum does decay significantly, suggesting a predominantly dynamical tunneling mechanism for IVR.  Two timescales in the case of the quantum decay, a fast component $t_{f} \sim 1000$  and a slower component $t_{s} \sim 35000$, are also evident from Figure~\ref{fig:sp_25_4_9}(a) which hints at the involvement of more than two states and multiple IVR pathways.  
The effect of increasing the coupling strengths on the survival probabilities are shown in Figure~\ref{fig:sp_25_4_9}(b)-(d) where the strength of the $(3,-2,0)$ resonance is kept relatively small in comparison to the $(2,-1,0)$ and the $(0,1,-2)$ resonances. The reason for this is that the junction-A is formed by the intersection of the latter two resonances and the ZOS of interest lies in the vicinity of this junction. This is also done to focus on the junction of interest and avoid certain complications that can arise due to higher order effects arising from the involvement of induced secondary resonances. Clearly, in contrast to the previous case, even a small increase in the coupling strengths leads to a fast classical decay as well, with timescales that are comparable to the quantum decay.   A key point to be noted here is that even for large values of the couplings, the quantum decay exhibits strong recurrences with $\sim 60\%$ of quantum probability cycling back to the initial ZOS. In contrast, the classical IVR dynamics seems to ``thermalize" very quickly and shows little recurrence. The survival probabilities on relatively short time scales, shown in Figure \ref{fig:sp_25_4_9}(e)-(h),  indicate that  the initial 
classical and quantum decay time scales become more and more comparable with increasing coupling strengths. 

Understanding the IVR dynamics implied by the results shown in Figure~\ref{fig:sp_25_4_9} in terms of the pathways in the QNS is hardly straightforward. However, we provide here a rationale based on comparing the classical and quantum dynamics on the Arnold web. We start by  analyzing the destination of the energy flowing out of the ZOS of interest, in terms of the time-dependent mode populations, by computing the cross survival probabilities. 
In Figure~\ref{fig:csp_25_4_9}(a)-(d) and (e)-(h), the classical and the quantum cross survival probabilities corresponding to the coupling ranges in Figure \ref{fig:sp_25_4_9} are shown. 
The various states shown in the quantum results of Figure~\ref{fig:csp_25_4_9}(e)-(h) account for $\gtrsim 80\%$ of the total probability. We compare (cf. Figure~\ref{fig:csp_25_4_9}(a)-(d)) the extent of energy flow into these states in the classical case as well since the primary focus here is to highlight the competition between classical and quantum IVR mechanisms. 
Moreover, the set of states that mix classically and quantum mechanically is not the same, but they are not mutually exclusive either. A complete and detailed analysis of the full set of states that mediate the classical and quantum IVR is not attempted in the present study.
To gain insights in terms of the role of the junction on the IVR pathways,  Figure~\ref{fig:csp_25_4_9}(i)-(l)  show the evolution of the classical dynamics of the ZOS of interest projected on to the Arnold web. We also indicate on the web the location of the states that dominate the quantum cross survival probabilities. In light of the results summarized in Figure~\ref{fig:csp_25_4_9}, we now briefly discuss the key features of the IVR dynamics for  the different coupling cases shown in Figure~\ref{fig:sp_25_4_9}.

\begin{figure*}
\begin{center}
\includegraphics[width=0.95\textwidth]{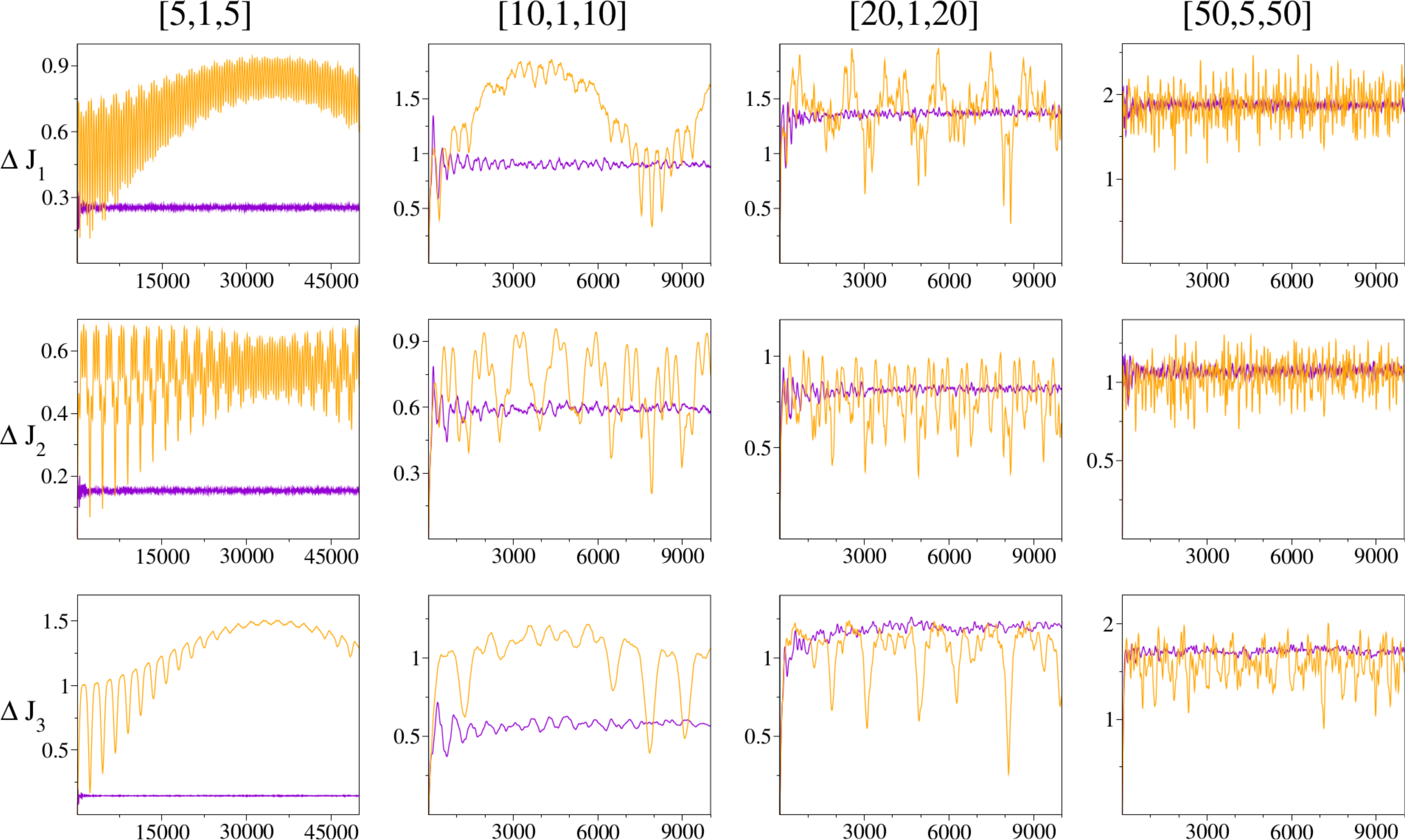}
\caption{The classical action drifts $\Delta J_{k}$ (purple) and the corresponding quantum number drifts $\Delta n_{k}$ (orange) as a function of time for various coupling strengths (indicated) are shown. See text for discussions.}
\label{fig:csp_25_4_9_adrift}
\end{center}
\end{figure*}

\begin{enumerate}
\item {\em Coupling strength}  $[5,1,5]$:  This case corresponds to the regime where the IVR pathways are purely quantum. This is immediately apparent from the survival probability of the ZOS in Figure~\ref{fig:sp_25_4_9}(a) and also from Figure~\ref{fig:csp_25_4_9}(a) which shows the cross survival probabilities. However,  most of the quantum probability does flow to the three  zeroth-order states  $|23,5,9\rangle$, $|25,3,11\rangle$, and $|23,3,13\rangle$. The location of these states on the Arnold web, as shown in Figure \ref{fig:csp_25_4_9}(i), is in the vicinity of the junction. Note that, based on the results shown in Figure~\ref{fig:csp_25_4_9}(e), the rate of energy flow into these three states is different. The timescale for energy transfer to $|23,5,9\rangle$ , $|25,3,11\rangle$, and $|23,3,13\rangle$ are $\sim 250, 1000$, and $35000$ respectively. These observations can be explained based on the dynamical tunneling mechanism.
The probability flow from $|25,4,9\rangle$ to $|23,5,9\rangle$ and $|25,3,11\rangle$ is due to RAT mediated by the  $(2,-1,0)$ and $(0,1,-2)$ resonances respectively. Again, the incomplete population transfer occurs due to the states not being symmetrically located with respect to the center of the appropriate resonances.  The  energy flow into the state $|23,3,13\rangle$, despite the lack of direct coupling to the initial ZOS by the primary resonances, comes from a RAT process that couples the state to $|25,3,11\rangle$.  The resonance that is responsible for this is $(2,0,-2)$, a higher order resonance that emanates from the junction of the  $(2,-1,0)$ and $(0,1,-2)$ resonances. Being a induced resonance, it has a significantly smaller width which leads to the long timescale of energy transfer seen in Figure~\ref{fig:csp_25_4_9}(e).   Although we do not explicitly show this here, it is possible\cite{ksjcp05,kspre05} to perturbatively remove the induced resonance and shut down the probability flow to $|23,3,13\rangle$, thereby validating our claim. Note, however, that at the present time we cannot comment about the interplay of the three RAT mechanisms and the extent of coherence between them. This requires further detailed study.

\item {\em Coupling strength} $[10,1,10]$: Here one already starts to see the coexistence of both classical and quantum IVR pathways. For instance, the location of the initial ZOS with respect to the junction, shown in Figure~\ref{fig:csp_25_4_9}(j), is such that classical IVR pathways do populate states $|23,5,9\rangle$ and $|25,3,11\rangle$. However, classically there is still no energy flow into  $|23,3,13\rangle$. Figure~\ref{fig:csp_25_4_9}(j) shows that for this  case there is an increase in the number of states that are involved in quantum mixing due to dynamical tunneling. Interestingly, a new quantum state $|21,5,11\rangle$ gets significantly populated, instead of the state $|23,3,13\rangle$ observed previously. Our analysis establishes that the states $|25,3,11\rangle$ and $|21,5,11\rangle$ are coupled by RAT involving the $(2,-1,0)$ resonance. Indeed, the coupling chain $|25,3,11\rangle \rightarrow |23, 4, 11\rangle \rightarrow |21,5,11\rangle$, along with the fact that very little probability builds up in the intermediate state, 
is indicative of a vibrational superexchange mechanism\cite{hutchinson84,stuchebrukhov93} which, as argued earlier\cite{ksirpc07}, is equivalent to dynamical tunneling.  The increased number of quantum states that mix due to dynamical tunneling leads to  enhanced, as compared to the classical case, probabilities observed in Figure~\ref{fig:csp_25_4_9}(f). Again, we defer a detailed analysis of the IVR mechanisms and the resulting dominant pathways to later studies.

\item {\em Coupling strengths} $[20,1,20]$ and $[50,5,50]$: In this regime the classical dynamics also mixes the quantum states of interest. 
In general, the reason for the classical mixing is either that the nonlinear resonances are wide enough to accommodate more than one state or, more importantly, the chaos 
around the junction connects the states.  From Figure~\ref{fig:csp_25_4_9}(k),(l) it is evident that
with increasing couplings all the relevant states around the junction are mixing both classically and quantum mechanically. Note that the coherent dynamical tunneling in the previous cases seems to have disappeared.
In particular, there is now a possibility of CAT leading to the faster timescales, and to some extent an incoherent, mixing seen in the quantum case.  However, establishing the extent to which CAT plays a role in the IVR process requires much more detailed studies and analysis. 
\end{enumerate}

As mentioned before, different routes to IVR for the classical and quantum dynamics couple different set of states and may lead to distinctly different extent of state space exploration. A measure for quantifying the extent and nature of exploration of the classical action space, equivalently the QNS, is the 
the evolution of the drift of the actions $\Delta J_{k}(t) \equiv {\sqrt{\langle J^{2}_{k} \rangle - \langle J_{k} \rangle ^2}}$ and the quantum analogue $\Delta n_{k}(t)$ with time.
In Figure \ref{fig:csp_25_4_9_adrift} we show the results of our computations for the ZOS $|25,4,9\rangle$ over the coupling range of interest.
For smaller couplings, the quantum $\Delta n$ dominates over the classical $\Delta J$  is due to dynamical tunneling.  The differences between the classical and quantum persists for the $[10,1,10]$ coupling case since dynamical tunneling continues to dominate the IVR process. In addition, as the classical mixing starts to contribute, 
one observes fast quantum oscillations since the  classically chaotic mixing contributes to the quantum IVR as well. However, with further increase in the coupling strengths, particularly for the case of $[50,5,50]$, Figure~\ref{fig:csp_25_4_9_adrift} shows that the quantum $\Delta n$ oscillates about the classical $\Delta J$, thus signaling a transition of the IVR dynamics from a purely quantum  to a mixed classical-quantum regime. Although we do not undertake any further analysis of the $P_{\bf n}(t)$ and $\Delta {\bf n}(t)$ in the present work, it would be useful to make connections with the scaling theory of IVR in order to assess the influence of the resonance junction on the QET.

\begin{figure*}[t]
\begin{center}
\includegraphics[width=0.90\textwidth]{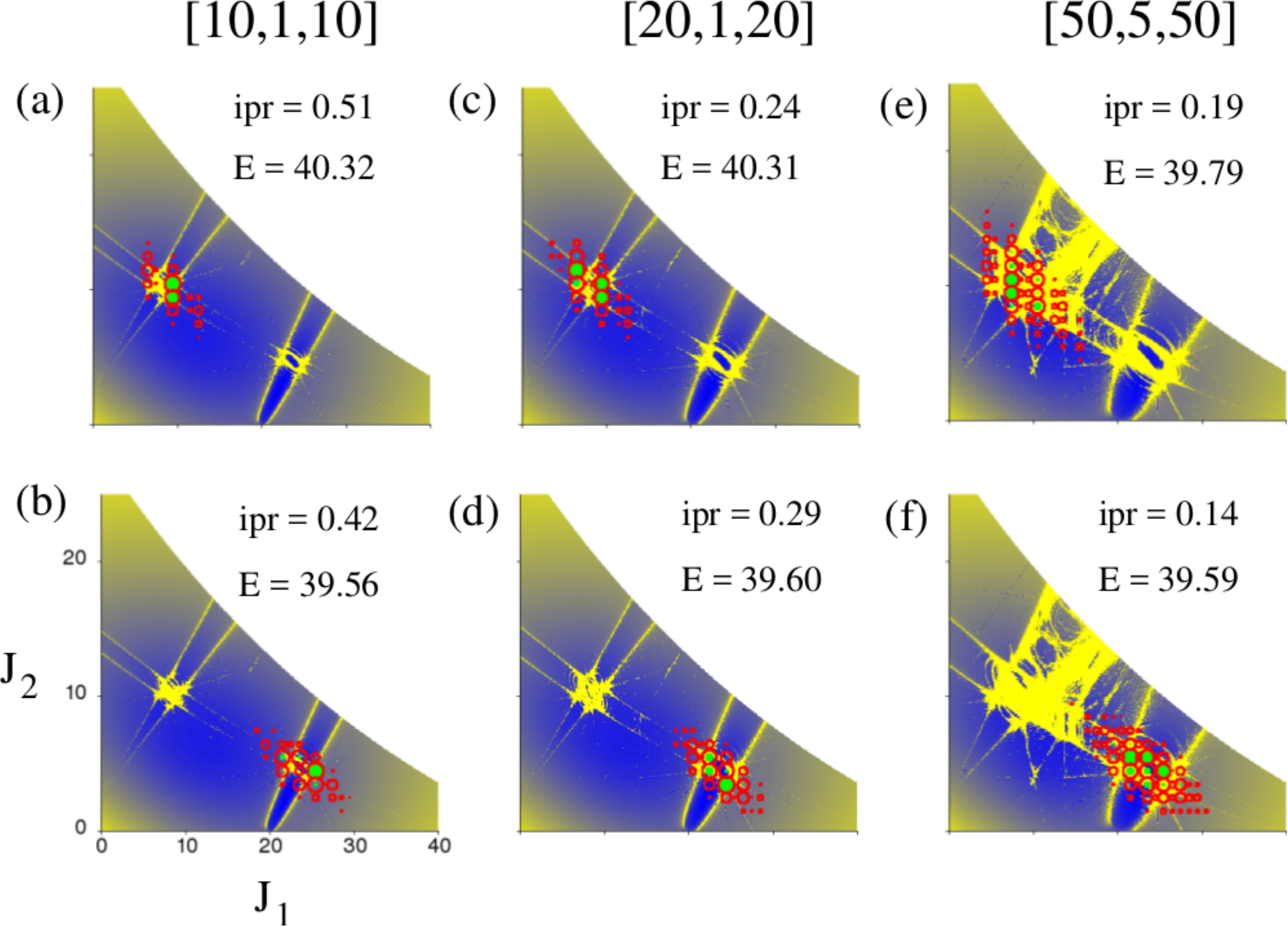}
\caption{Example  eigenstates for different coupling strengths (indicated) that are influenced by the resonance junctions. For every eigenstate $|\psi_{i}\rangle$ the eigenvalue $E$ and the corresponding inverse participation ratio (ipr), computed as $\sum_{{\bf n}} |\langle {\bf n}|\psi_{i} \rangle|^{4}$, are also indicated. States that contribute dominantly to an eigenstate are indicated in green.}
\label{fig:eigenstates}
\end{center}
\end{figure*}

\subsection{Nature of the eigenstates: thermalization versus dynamical tunneling} 
\label{eigenstate_delocalization}

In the previous sections it was shown  that  the initial states near resonance junctions exhibit a fairly rich and complex dynamics. Consequently, it is natural to expect that the classical and quantum mixing near the junctions must leads to the 
formation of a class of interesting quantum eigenstates. In this section we give examples of few such eigenstates and reserve a more detailed analysis for later work. Since we have mapped the Arnold web on the classical action space, and given the results in Figure~\ref{fig:web_qumweb}, it is convenient to project the eigenstates  on the QNS for a direct comparison.  Such a representation of the   
eigenstates is expected to reveal the effect of the classical Arnold web structures on the quantum eigenstates, which encode the long time IVR dynamics of the system. However,  it is difficult to visualize the eigenstates in the three dimensional QNS. Hence, a reduced\cite{ccmjstat92} QNS representation is constructed as follows. We define a reduced density 
for an eigenstate $|\psi_{k}\rangle$ as
\begin{equation}
 \rho'_{k}(n_1,n_2) = \sum_{n_3} | \langle {\bf n} | \psi_{k} \rangle |^2 \equiv \sum_{n_3} |c_{{\bf n} k}|^{2}
\end{equation}
and associate $\rho'_{k}(n_1,n_2) \equiv |\tilde{c}_{n_1,n_2;k}|^{2}$.
For ease of visualization,  the reduced density is scaled as follows. 
 If $|\tilde{c}_{n_1,n_2};k|^2 > 10^{-6}$ then the coefficients are scaled as $|c'_{n_1,n_2;k}|^2 = \log(|\tilde{c}_{n_1,n_2;k}|^2 \times 10^6)/6$, otherwise we take $|c'_{n_1,n_2;k}|^2 = 0$. We thus represent the eigenstates in the projected QNS by plotting  
 $|c'_{n_1,n_2;k}|^2$ at every point $(n_1,n_2)$. Note that this choice of scaling is arbitrary and a different cutoff for $|c'_{n_1,n_2;k}|^2$ would reveal even finer details of the tails of the eigenstates.

\begin{figure}[!t]
\begin{center}
\includegraphics[width=1\linewidth]{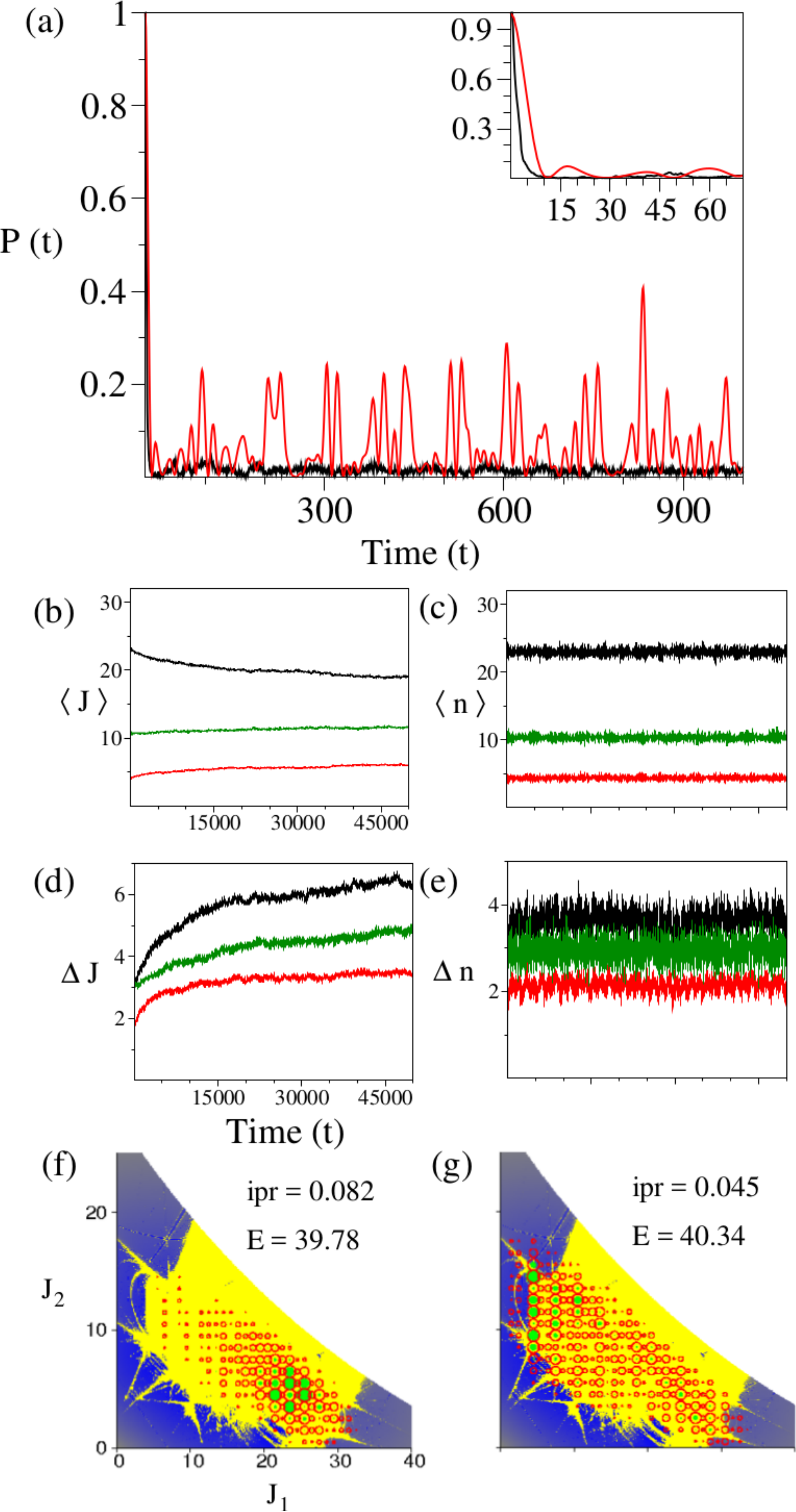}
\caption{(a) The classical (black) and quantum (red) survival probabilities of an initial ZOS (25,4,9) for the coupling 
$[200,10,200]$. In inset, the initial time decay of both classical and quantum. (b) and (c) are the 
average of the classical actions and quantum numbers respectively. (d) and (e) are the 
classical $\Delta J$ and quantum $\Delta n$ respectively. (f) and (g) are two eigenstates projected on 
the Arnold web.}
\label{fig:sp_200_10_200}
\end{center}
\end{figure}

In Figure \ref{fig:eigenstates}, few example eigenstates near the resonance junctions are shown. Note that in the figures, the reduced density is represented as red circles with radii proportional to $|c'_{n_1,n_2;k}|^2$ and the green color indicates ZOS with dominant contribution to the specific eigenstate. It is clear that the eigenstates are more delocalized  for larger values of the coupling strengths. An interesting feature of the eigenstates is the delocalization  along resonances. For instance, the eignstate in Figure~\ref{fig:eigenstates}(e) exhibits partial delocalization along the $(0,1,-2)$ resonance channel. This delocalization along a resonance channel is reminiscent, as also noted by Martens\cite{ccmjstat92}, of the phenomenon of Arnold diffusion. It is tempting to make this association since Arnold diffusion is expected to be an extremely long time process and hence the best chance to observe its quantum counterpart is in the eigenstates. At the same time,  one expects this mechanism of transport on the web to be localized due to quantum effects and hence associated with partially delocalized eigenstates. 
Nevertheless, as evident from the results in Figure \ref{fig:csp_25_4_9_adrift}, the classical action never access the regions of QNS corresponding to  the tails of the eigenstates. Thus, the spreading along the resonance channels must be predominantly due to dynamical tunneling.  This is confirmed further by our finding of other eigenstates that are nearly degenerate with those shown in Figure~\ref{fig:eigenstates}. It remains to be seen if tuning the effective $\hbar$ down may result in eigenstates whose delocalization along resonance channels is due to a dominant classical mechanism.

Finally, as an example case, In Figure~\ref{fig:sp_200_10_200} we show the dynamics for a very large coupling strength $[200,10,200]$. In this regime all three resonances overlap and lead to widespread chaos. The IVR dynamics of the ZOS $|25,4,9\rangle$
is shown in Figure~\ref{fig:sp_200_10_200}(a) and it is seen that both classical and quantum survival probabilities decay very fast with very similar short  time decay profiles. Despite this, the classical dynamics thermalizes very rapidly but quantum recurrences are still present and prevent thermalization. 

It is interesting to note that the  expectations of the classical 
action and quantum number as a function of time, shown in  Figure~\ref{fig:sp_200_10_200}(b)-(c), are not significantly different.  However, the time dependence of the action and quantum number drift  shown in Figure \ref{fig:sp_200_10_200}(d)-(e) indicate that  the extent of the state space explored is very different for the classical and quantum dynamics. 
 The quantum dynamics in this case is more localized than the classical dynamics. These results need further analysis in terms of the phenomenon of thick layer diffusion and the Shuryak criterion\cite{shuryak76} for quantum delocalization. Briefly, thick layer diffusion refers to the diffusion due to the extensive chaos that is generated near resonance junctions due to significant overlap of the resonances. The Shuryak criterion provides a bound on the effective $\hbar$ value that is necessary to observe significant quantum diffusion, which in turn  is related to the extent of delocalization and density of the quantum eigenstates in the chaotic layer. A theoretical understanding of the competition between classical and quantum transport in this regime can be obtained from the stochastic pump model. However, we note that very few studies have addressed this issue in the regime that is pertinent to Figure~\ref{fig:sp_200_10_200}. The first one by Leitner and Wolynes\cite{leitwoly97} provides a better estimate then the original one by Shuryak and predicts that the localization length of the quantum states should  exhibit  a polynomial rather than an exponential scaling.  Interestingly, a later work by Malyshev and Chizova\cite{Malyshev2010}  finds that the quantum diffusion, under certain conditions, is insensitive to the Shuryak constraint. Clearly, our understanding of the classical versus quantum IVR dynamics in this physically relevant regime is still rather limited.

\section{Concluding remarks} 
\label{conclusion}

In this work we have explored the competition and correspondence between classical and quantum IVR dynamics for a  model Hamiltonian with three degrees of freedom. 
This study  can be thought of as a natural next step to the detailed studies performed nearly three decades ago by Davis and Heller\cite{davisheller84} on model systems with two degrees of freedom. 
Our results highlight the complex energy flow dynamics that occurs for initial states that are located in the vicinity of the resonance junctions on the Arnold web. 
Near a junction, presence of several resonances and chaos provides numerous competing IVR pathways that preclude a full mechanistic understanding of IVR at this time. 
However, our results clearly establish the transition of the IVR dynamics from a coherent dynamical tunneling dominated regime to a incoherent chaos-assisted regime. As found in the earlier study\cite{davisheller84}, we observe restricted exploration of the quantum state space even for resonant coupling values  that lead to extensive chaos. Interestingly, we find that the quantum dilution factors  mirror the classical Arnold web structure over a wide range of the resonant coupling strengths. This observation correlates well with the central point of LRMT that the local density of states, and not the total density of states, is the critical quantity that determines the transition to facile IVR\cite{bigwood1998}. We have also shown a few example eigenstates, projected on to the Arnold web, which bring out the importance of dynamical tunneling to eigenstate delocalization. The analysis presented here can be utilised to identify the manifestation of resonance junctions in the corresponding quantum IVR dynamics.

Regarding the question of whether there exists a classical limit to QET, the current work must be considered as an preliminary attempt to  determine the extent to which quantum dynamical tunneling persists, and competes with classically chaotic IVR pathways, as one makes the transition from the KAM to the Chirikov regime. For instance, it is in the intermediate Nekhoroshev regime, as seen in the results presented here,  that a clear distinction between classical and quantum IVR  emerges due to dynamical tunneling. However,  beyond the Nekhoroshev regime, the problem at hand is complicated by the fact that dynamical tunneling is not only influenced by the various resonances but also by the classical chaos. The competing contributions, from enhanced tunneling due to CAT and quantum localization due to a finite $\hbar$, to the various terms in QET need to be carefully dissected. Thus, there still remain a number of issues, not addressed in the present study, and we list a few of them below.

 A complete understanding of the classical-quantum correspondence requires\cite{pittmanheller16} variations in the effective $\hbar$ and the influence on the IVR dynamics and dynamical tunneling. This is particularly important to establish whether or not chaos-assisted tunneling near the junctions needs to be considered. It is crucial to do a more detailed analysis\cite{haller95,efthy2008,cincotta2014} in terms of putting bounds on the extent of action drift that one expects classically in various regimes. This would then provide a cleaner basis for quantifying the extent of eigenstate delocalization due to dynamical tunneling alone. In the parameter regimes considered herein the two main junctions were far apart in the state space. What happens if they are closer or there are several more dominant junctions? This situation is expected to arise in realistic systems and currently there is very little known about the IVR dynamics in these cases. As a final comment, note that the eigenstate shown in Figure~\ref{fig:sp_200_10_200}(g) seems to involve dominant contributions from both junctions. Understanding the mechanisms that give rise to such classes of states is expected to shed light on the validity of ETH and the possibility of novel MBL phases in molecular systems. Some of the issues highlighted here are currently under investigation in our group.

\begin{acknowledgements}

Sourav Karmakar thanks the University Grants Commission (UGC), India for a doctoral fellowship. This work is supported by the Science and Engineering Research Board (SERB) of India under the grant 
EMR/2016/0062456. We acknowledge the IIT Kanpur High Performance Computing facility for computing resources.

\end{acknowledgements}


\end{document}